\documentclass[twocolumn,aps,superscriptaddress,preprintnumbers]{revtex4}
\usepackage{graphicx}
\usepackage{amssymb}
\usepackage{epstopdf}
\usepackage{braket}

\usepackage{bm} 
\usepackage{longtable}
\usepackage{multirow}

\usepackage{hyperref}
\hypersetup{
    colorlinks=true,       
    linkcolor=blue,        
    citecolor=blue,        
    filecolor=blue,        
    urlcolor=blue,         
    runcolor=blue
}

\usepackage{xcolor}

\begin{document}

\title{Dark Vibronic Polaritons and the Spectroscopy of Organic Microcavities}
\author{Felipe Herrera}
\email{felipe.herrera.u@usach.cl}
\affiliation{Department of Physics, Universidad de Santiago de Chile, Avenida Ecuador 3943, Santiago, Chile}
\author{Frank C. Spano}
\email{spano@temple.edu}
\affiliation{Department of Chemistry, Temple University, Philadelphia, Pennsylvania 19122, USA}

\date{\today}                                           

\begin{abstract}

Organic microcavities are photonic nanostructures that strongly confine the electromagnetic field, allowing exotic quantum regimes of light-matter interaction with disordered organic semiconductors. The unambiguous interpretation of the spectra of organic microcavities has been a long-standing challenge due to several competing effects involving electrons, vibrations and cavity photons. Here we present a theoretical framework that is able to describe the main spectroscopic features of organic microcavities consistently. We introduce a class of light-matter excitations called dark vibronic polaritons,  which strongly emit but only weakly absorb light in the same frequency region of the bare electronic transition. Successful comparison with experimental data demonstrates the applicability of our theory. The proposed microscopic understanding of organic microcavities paves the way for the development of optoelectronic devices enhanced by quantum optics.

\end{abstract}
\maketitle

Optical microcavities containing organic semiconductor materials can confine the electromagnetic field within tens of nanometers at room temperature. Such small field volumes allow current experiments to reach the regimes of strong \cite{Lidzey1998,Tischler2007,Torma2015} and ultrastrong \cite{Schwartz2011,Kena-Cohen2013,Gambino2015} coupling of cavity quantum electrodynamics (QED) with organic matter. 
The quantum optical properties of these organic microcavities can be analogous to atomic cavities \cite{Mabuchi2002} or superconducting resonators \cite{You2011}, but molecular systems are unique because they can undergo a variety of chemical processes involving electronic and vibrational degrees of freedom, while simultaneously interacting with a confined electromagnetic vacuum. This versatility has stimulated the development of nanoscale molecular devices whose electronic transport properties \cite{Orgiu2015} or chemical reactivity \cite{Hutchison:2012,Herrera2016} can be manipulated using quantum optics. 

 Organic semiconductor materials are known to have strong coupling of electronic transitions with intramolecular vibrations (phonons), which  determines the photophysics \cite{Spano2010} and transport properties \cite{Coropceanu2007} of organic materials. In microcavities, the electron-phonon coupling can be comparable in strength with the so-called vacuum Rabi frequency, which quantifies the strength of light-matter interaction. The resulting interplay between electrons, vibrations and cavity photons has for years obscured the interpretation of  absorption and emission data. For instance, early experiments \cite{Hobson2002} showed the emergence of strong  emission lines but negligible absorption in the frequency region near the bare molecular resonance. Such results seem unexpected because  reciprocity dictates that a strong emitter should also be a good absorber \cite{May-Kuhn}.  Several experiments have since then reproduced these observations \cite{Coles2011,Virgili2011,Schwartz2013}, and identified further spectral features of organic microcavities \cite{George2015-farad} that cannot be consistently explained by existing quasi-particle theories \cite{Agranovich2003,Litinskaya2004,Litinskaya2006,Cwik2016}. These unusual observations have been made on systems where the Rabi frequency exceeds the typical frequency of the vinyl stretching mode, characteristic of conjugated organic molecules \cite{Spano2010},  below the regime of polaron decoupling \cite{Herrera2016}. Our goal is to understand the polariton structure in this intermediate coupling regime, with a focus on the spectral region in the middle of the conventional lower and upper polariton doublet, where quasi-particle theories predict the existence of an incoherent exciton reservoir \cite{Agranovich2003,Litinskaya2004,Litinskaya2006,Michetti2008,Mazza2009,Cwik2016} . 
 
In order to achieve this goal, we develop here a theoretical framework that is able to account for several unexplained features of the optical spectra of organic microcavities consistently. The theory is based on the Holstein-Tavis-Cummings (HTC) model \cite{Cwik2014,Spano2015,Herrera2016,Zeb2016,Wu2016},  and a Lindblad approach for the dissipative dynamics of organic polaritons. We provide a microscopic interpretation of organic microcavity spectroscopy in terms of {\it dark vibronic polaritons}, collective light-matter states that weakly absorb but strongly emit radiation. We show that dark vibronic polaritons can emerge by two possible mechanisms: (\emph{a}) the destructive interference  between diabatic polariton states associated with different vibronic transitions, and (\emph{b}) the admixture of diabatic two-particle states \cite{Spano2010} with dark excitonic material states. 


We describe an ensemble of $N$ organic emitters inside an optical cavity  
 with the Holstein-Tavis-Cummings (HTC) model \cite{Cwik2014,Spano2015,Herrera2016,Zeb2016,Wu2016}, which can be written as $\hat{\mathcal{H}}=\hat H_{\rm C}+\hat H_{\rm M}+\hat V_a$.  $\hat H_{\rm C}=\omega_c \,\hat a^\dagger \hat a$ is the free cavity Hamiltonian, the molecular degrees of freedom are described by the Holstein model \cite{Holstein:1959,Spano2010}
 \begin{equation}\label{eq:Holstein}
\hat H_{\rm M}= \omega_{\rm v}\sum_{n=1}^N\hat b_n^\dagger \hat b_n+\sum_{n=1}^N\left[\omega_{e}+\omega_{\rm v}\lambda(\hat b_n+\hat b_n^\dagger)\right]\ket{e_n}\bra{e_n},
\end{equation}
and the cavity-matter coupling by a Tavis-Cummings \cite{Garraway2011} term 
\begin{equation}\label{eq:Tavis-Cummings}
\hat V_a = (\Omega/2)\sum_{n=1}^N(\ket{g_n}\bra{e_n}\hat a^\dagger +\ket{e_n}\bra{g_n}\hat a).
\end{equation}
The vertical Frank-Condon transition frequency is 
$\omega_e =\omega_{00}+ \omega_{\rm v}\lambda^2$, where $\omega_{00}$ is the frequency of the zero-phonon (0-0) transition, $\omega_{\rm v}\approx 0.15-0.18$ eV is the intramolecular vibrational frequency,  and $\lambda^2\sim 0.1-1.2$ is the Huang-Rhys factor. 
 Operator $\hat b_n$ annihilates an intramolecular phonon on the $n$-th chomophore, $\hat a$  annihilates a cavity photon,  and $\Omega$ is the single-particle vacuum Rabi frequency. 
Throughout this work,  we assume a cavity detuning $\Delta \equiv \omega_{00}-\omega_{c}=0$ at normal incidence. 

We are interested in the eigenstates of the Holstein-Tavis-Cummings (HTC) Hamiltonian $\hat{\mathcal{H}}$ whose energy equals the bare molecular transition, since unexpectedly strong cavity emission is observed in this frequency  region \cite{Hobson2002,Coles2011,Virgili2011,Schwartz2013}. In the rotating frame of the resonant cavity mode, a polariton state at the bare molecular frequency satisfies $\hat{\mathcal{H}}\ket{X}=0$. It is straightforward to show \cite{Herrera2016-PRA}, that a vanishing eigenvalue implies that the transition dipole moment $\mu_{XG}\equiv\langle G|\hat \mu|X \rangle$ also vanishes, for $\ket{G}$ being the absolute ground state of the cavity (no vibronic, vibrational or cavity excitations), and $\hat \mu $ the dipole operator. In general, we refer to any HTC eigenstate $\ket{\epsilon_j}$ with vanishing transition dipole with the absolute ground state ($\mu_{jG} = 0$), as a {\it dark vibronic polariton}, since it is effectively invisible in room temperature absorption experiments that probe material dipole transitions directly, for instance, using bound modes of the nanostructure \cite{Matterson2001,Torma2015,Herrera2016-PRA}. 

The vanishing dipole strength of the zero-energy dark vibronic polariton $\ket{X}$ can be understood as a destructive interference effect. For $N=1$, we introduce the diabatic polariton states
\begin{equation}\label{eq:diabatic polaritons}
\ket{\nu_\pm}=\frac{1}{\sqrt{2}}\left(	\ket{e,\tilde \nu,0_c}  \pm \ket{g,\nu,1_c}\right),
\end{equation}
where $\ket{g,\nu,1_c}$ represents a molecule in its ground state $\ket{g}$ and vibrational eigenstate $\ket{\nu}$ with one cavity photon, and $\ket{e,\tilde \nu,0_c}$ represents a molecule in the excited state $\ket{e}$, and vibrational eigenstate $\ket{\tilde \nu}$ in the cavity vacuum. The tilde overstrike indicates that the vibrational eigenstate has $\tilde\nu$ vibrational quanta in the excited state harmonic nuclear potential, whose minimum is shifted relative to the ground state potential minimum by a quantity $\lambda$ along a dimensionless vibrational coordinate \cite{May-Kuhn,Spano2010}. In the diabatic basis of Eq. (\ref{eq:diabatic polaritons}), the number of vibrational excitations in the ground and excited state potentials is conserved.

The light-matter coupling term of the HTC model [Eq. (\ref{eq:Tavis-Cummings})] breaks this symmetry and admixes vibronic polariton states with different number of vibrational quanta. In a truncated diabatic polariton basis [Eq. (\ref{eq:diabatic polaritons})], we can expand this state as $\ket{X}\approx c_0\ket{0_+}-c_1\ket{1_-}-c_2\ket{2_+}$, giving
$\mu_{XG} \approx \mu \left(c_0\langle 0|\tilde 0\rangle -c_1\langle 0 |\tilde 1\rangle-c_2\langle 0 |\tilde 2\rangle\right )$, 
where $\mu$ is the molecular transition dipole moment, $c_{\nu}> c_{\nu+1}>0$, and $\lambda>0$. In other words, the destructive interference between the Frank-Condon factors of bare vibronic transitions leads to a vanishing transition dipole. However, state $\ket{X}$ can still emit light at frequency $\omega=-\nu\omega_{\rm v} $, with $\nu\geq 0$, upon photon leakage through the cavity mirrors.

For a single molecule, the zero-energy dark vibronic polariton $\ket{X}$ forms at the critical coupling $\Omega=1.7\,\omega_{\rm v}$, i.e., $\hat{\mathcal{H}}\ket{X}=0$. For $N\geq 2$, the critical Rabi coupling is in the range $\sqrt{N}\Omega/\omega_{\rm v}\approx 1.8-2.5$ for $\lambda^2\sim 0.1-1.0$ \cite{Herrera2016-PRA}. For large ensembles, the HTC model can support multiple vibronic polariton eigenstates in the vicinity of the bare molecular frequency with small or vanishing transition dipole to the absolute ground state ($\mu_{jG}\approx 0$), mainly due to destructive interference effects like the one discussed above.  Such states are non-degenerate and we refer to them as dark vibronic polaritons of the $X$ type.

 Two-particle (TP) material states become important to describe the spectra of molecular ensembles ($N\geq 2$) \cite{Spano2015}. These states are represented by  $\ket{e_n,\tilde \nu_n,g_m,\nu_m,0_c}$, where $n$ labels the location of a vibronic excitation with $\tilde \nu$ displaced vibrational quanta, while molecule $m$ is in a vibrational eigenstate with $\nu\geq 1$ vibrational quanta \cite{Herrera2016-PRA}. Two-particle {\it material} states give rise, on resonance, to the two-particle {\it polariton} states   \cite{Spano2015,Herrera2016-PRA} 
\begin{equation}\label{eq:TP polariton}
\ket{{\rm P}^\pm_{ \nu \tilde\nu'}\beta}=\frac{1}{\sqrt{2}}\left( \ket{\alpha_0\beta,\tilde \nu'\nu,0_c}\pm \ket{\beta,\nu,1_c}\right)
\end{equation}
where the first term is the permutation-symmetric two-particle material state 
\begin{equation}\label{eq:TP particle}
\ket{\alpha_0\beta,\tilde \nu'\nu,0_c}=\sum_{m=1}^N\sum_{n\neq m}^N\frac{c_{\beta,m}}{\sqrt{N-1}}\ket{e_n,\tilde \nu'_n,g_m,\nu_m,0_c},
\end{equation}
and the photonic component is given by
\begin{equation}\label{eq:photonic}
\ket{\beta,\nu,1_c} = \sum_{m=1}^Nc_{\beta m}\ket{g_m,\nu_m,1_c}
\end{equation}
which is a product of a collective vibrational excitation and a single-photon state. The quantum number  $\beta$ in Eq. (\ref{eq:TP polariton}) quantifies the symmetry under particle permutation of the material states, encoded in the amplitudes $c_{\beta m}$. We associate the value $\beta=\alpha_0$ to the totally-symmetric superposition, i.e., $c_{\alpha_0,m} = 1/\sqrt{N}$, for all $m$. In the language of molecular aggregates \cite{Spano2010}, this would correspond to a state with zero momentum. 
For $\omega_c = \omega_{00}$, polaritons in Eq. (\ref{eq:TP polariton}) have the $N$-fold degenerate diabatic energies  
\begin{equation}\label{eq:Enu}
E_{\tilde \nu, \nu}^{\pm} = (\tilde \nu+\nu)\omega_{\rm v}\pm \sqrt{N-1}|\langle 0|\tilde \nu \rangle|\Omega/2,
\end{equation} 
which form a polariton doublet for small Rabi couplings $\sqrt{N}\Omega/\omega_{\rm v}\ll 1$ \cite{Herrera2016-PRA}.

For larger Rabi frequencies \mbox{$\sqrt{N}\Omega/\omega_{\rm v}\approx 2$}, the two-particle diabatic polaritons given by Eq. (\ref{eq:TP polariton}) that are permutation-symmetric ($\beta=\alpha_0$) can admix with conventional (single-particle) vibronic polariton states  \cite{Herrera2016,Herrera2016-PRA}
\begin{equation}\label{eq:SP polariton}
\ket{\alpha_0, \tilde \nu,\pm} = \frac{1}{\sqrt{2}}\left(\ket{\alpha_0,\tilde \nu,0_c}\pm\ket{g_10_1,g_20_2\ldots,g_N0_N,1_c}\right), 
\end{equation}
where $\ket{\alpha_0,\tilde \nu,0_c} = \sum_n \ket{e_n\tilde \nu_n,0_c}/\sqrt{N}$ for $\tilde \nu\geq 0$, to form a set of {\it non-degenerate} states that we designate as the dark vibronic polaritons of the $X$ type. Such polaritons are have vanishing or negligible transition dipole moment with the absolute ground state of the system $\ket{G}$, but nevertheless contribute to the photoluminescence spectra. 

 \begin{figure}[t]
\includegraphics[width=0.42\textwidth]{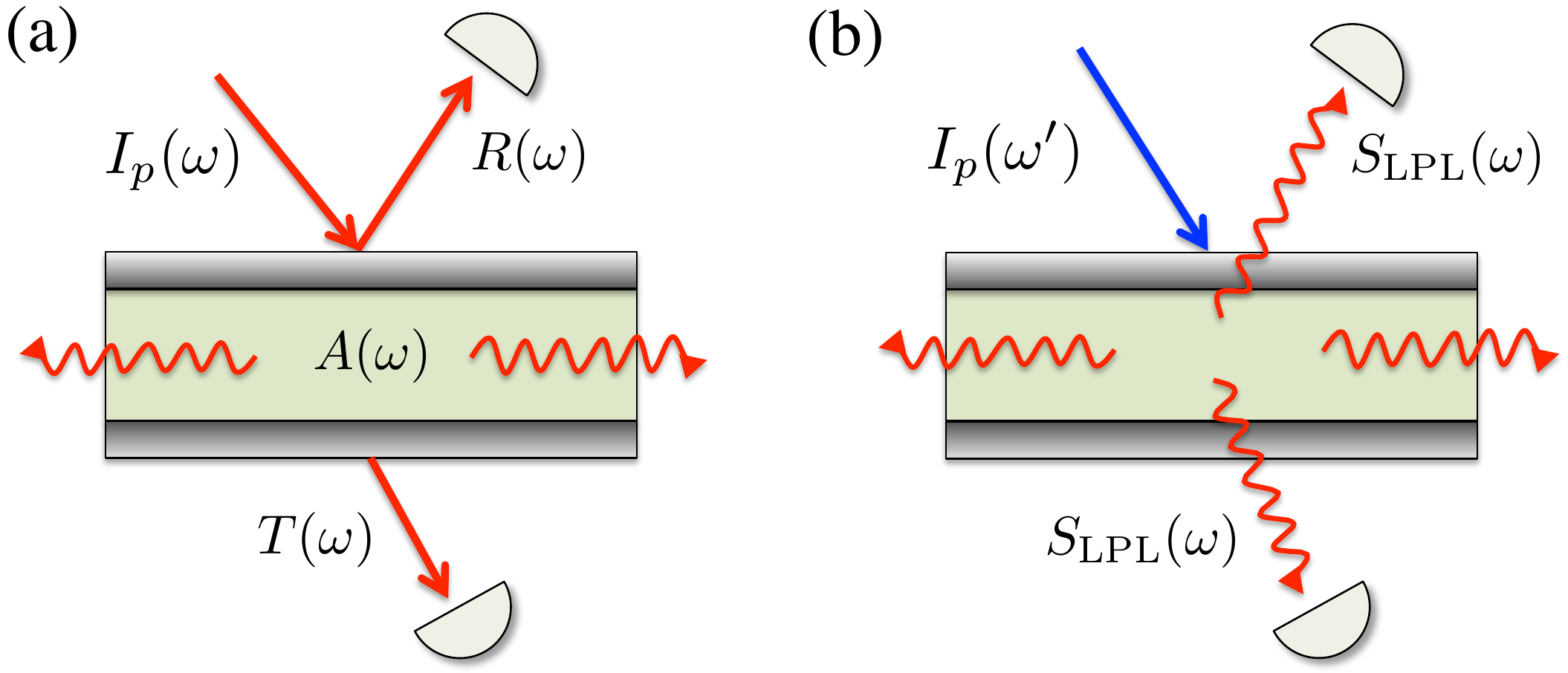}
\caption{Spectral signals. (a) Reflection $R$, transmission $T$, and absorption $A = 1-R-T$ of an external pump $I_p(\omega)$. Absorption is due to spontaneous emission into bound modes of the microcavity.  (b) Leakage photoluminescence $S_{\rm LPL}(\omega)$ following weak laser pumping at frequency $\omega'>\omega$. Bound photoluminescence \cite{Spano2015} is shown as horizontal wavy arrows. }
\label{fig:definitions}
\end{figure}

While $X$-type dark vibronic polaritons involve the superposition of states that are totally-symmetric with respect to permutation, the HTC model also allows the admixture of single and two-particle states that are not totally-symmetric ($\beta\neq \alpha_0$). We refer to the resulting eigenstates as dark vibronic polaritons of the $Y$ type. These are invisible in both conventional (through-mirror) and bound mode absorption \cite{Herrera2016-PRA}. As a specific example of this mixing between non-symmetric states, we can consider the {\it material} states $\ket{\beta\neq \alpha_0,\tilde 0,0_c}$, commonly known as dark exciton states \cite{Herrera2016,Gonzalez2016}, which admix with the non-symmetric two-particle diabatic polariton state $\ket{P^{-}_{1\tilde 0},\beta\neq 0}$, for each of the $N-1$ values of $\beta$ [see Eq. (\ref{eq:TP polariton})]. For Rabi couplings \mbox{$\sqrt{N}\Omega/\omega_{\rm v}\approx 2$}, the states involved in the mixture  are nearby in energy and generate a manifold of $N-1$ dark vibronic polariton states of the $Y$ type near the bare molecular resonance  ($\omega_j/\omega_{\rm v}\sim 0.1$), which can be approximately written as 
 \begin{equation} \label{eq:Y state}
 \ket{Y_{j}}\approx a_j\ket{\beta,\tilde 0,0_c}+b_j\ket{P^{-}_{1\tilde 0},\beta}, 
 \end{equation}
with  $|a_j|\sim |b_j|$. This set of $Y$-type vibronic polaritons carry a photonic component with one quantum of vibration. Light emitted by state $\ket{Y_j}$ through photon leakage therefore has frequency $\omega= \omega_j-\omega_{\rm v}$, which is near the lower polariton frequency for the Rabi frequencies considered. In addition to the states given by Eq. (\ref{eq:Y state}), there  are also multiple dark vibronic polaritons of the $Y$ type near the conventional upper polariton frequency $(\omega_j\approx \omega_{\rm v})$, having photonic components with one or more vibrational excitations. Such states can emit light near the bare molecular resonance frequency ($\omega\approx 0$) upon photon leakage. As we show below, the structure of dark vibronic polaritons of the $Y$ type greatly influences the photoluminescence spectra of organic cavities.

In order to understanding specific spectral features of dark vibronic polaritons of the $X$ and $Y$ types,  we focus on the two most commonly measured signals: absorption $A(\omega)$, and leakage photoluminescence $S_{\rm LPL}(\omega)$. These are illustrated in Fig. \ref{fig:definitions}.  In general, there are two types of absorption experiments, depending on whether a  resonant laser pump drives the cavity through the mirrors \cite{George2015-farad}, as in Fig. \ref{fig:definitions}, or by directly populating a bound mode of the nanostructure \cite{Matterson2001}. We refer to the latter as bound absorption. Similarly, once a polariton is populated, it can lose its energy by photon leakage through the mirrors, generating the leakage photoluminescnce (LPL) signal, or by fluorescence into bound modes \cite{Torma2015,Spano2015}. The latter process is responsible for resonant pump attenuation and thus determines the conventional absorption spectra by the relation $A=1-R-T$. 

\begin{figure*}
\includegraphics[width=\textwidth]{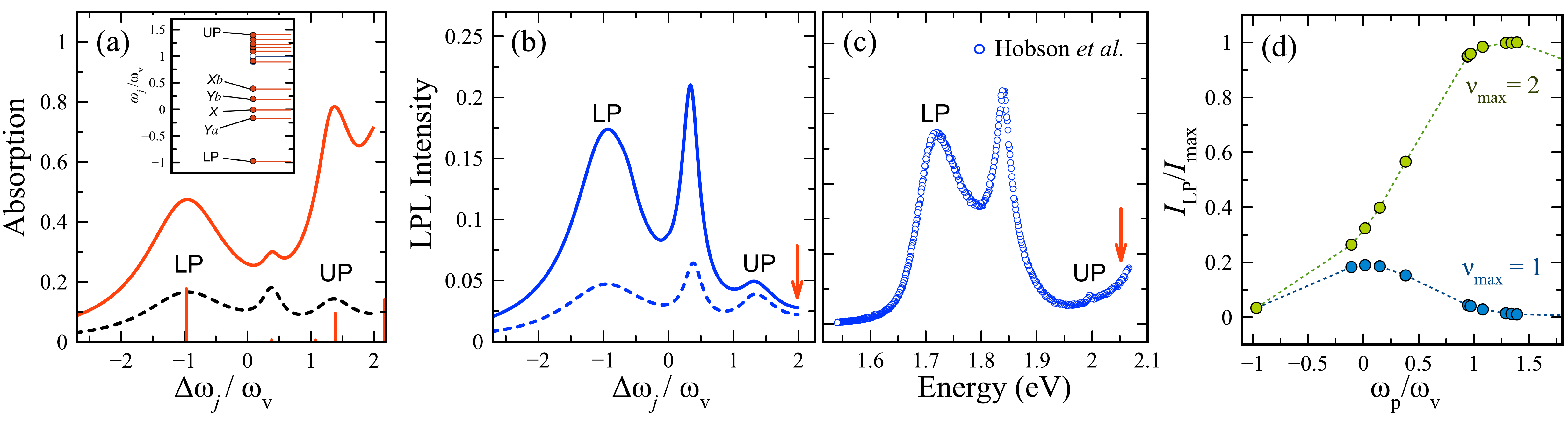}
\caption{Model microcavity spectra for $N = 20$ emitters and comparison with experiment. (a)  Absorption spectra $A  = 1-R-T$ (dashed line) and bound mode absorption (solid line), indicating the lower (LP) and upper polariton (UP) peaks. Vertical bars indicate the position and relative oscillator strength of the bound absorption lines. (b) Leakage photoluminescence (LPL) spectra including dissipative transitions that project the material onto the absolute ground state (dashed line), or to states in the ground manifold with up to one quantum of vibration (solid line). The red arrow indicates the highest energy  polariton  included in the simulation. (c) Experimental LPL spectra for an ensemble of cyanine dye J-aggregates in a microcavity, reprinted from Ref. \cite{Hobson2002}. The red arrow indicates the laser pump energy. (d) Relative emission intensity at the frequency of the lower polariton $I_{\rm LP}$, for a Gaussian polariton population centered at $\omega_{\rm p}$ with standard deviation $\sigma_{\rm p}/\omega_{\rm v} = 0.5$. Transitions to the ground state with up to $\nu_{\rm max}$ vibrational quanta are included. We use  $\lambda^2 = 1$, $\sqrt{N}\Omega/\omega_{\rm v}=2.4 $, $\kappa/\omega_{\rm v}=0.9$, and $N\gamma_e/\omega_{\rm v}=3.0$, where $\omega_{\rm v}$ is the vibrational frequency. The cavity is assumed resonant with the molecular transition at normal incidence}
\label{fig:experiment}
\end{figure*}

We derive, for the first time, expressions for the conventional absorption and leakage PL spectra in organic microcavities, using a standard cavity QED approach \cite{Herrera2016-PRA}. The resulting expressions depend on the generalized spectrum of fluctuations $S(\omega) = \sum_j\rho_j\, S_{\hat O}^{(j)}(\omega)$, where
\begin{equation}\label{eq:S omega}
S_{\hat O}^{(j)}(\omega) = \sum_{i}\,|\langle\epsilon_i| \hat O| \epsilon_j\rangle |^2 \frac{\kappa_{j}}{(\omega-\omega_{ji})^2+\kappa_{j}^2},
\end{equation}
is a state-dependent lineshape function, and $\kappa_j\approx \Gamma_j/2$ is the decay rate of the polariton coherence. 
 States $\ket{\epsilon_j}$ and $\ket{\epsilon_i}$ are eigenstates of the Holstein-Tavis-Cummings model in the one-polariton and ground state manifolds, respectively.  $\rho_j$ is the stationary population of the $j$-th polariton eigenstate, $\omega_{ji}>0$ is the frequency of the transition $\ket{\epsilon_j}\leftrightarrow \ket{\epsilon_i}$ and $\Gamma_j = \kappa\sum_i|\bra{\epsilon_i}\hat a\ket{\epsilon_j}|^2+N\gamma_e\sum_i|\bra{\epsilon_i}\hat J_-\ket{\epsilon_j}|^2$ is the polariton decay rate.
 $\kappa$ is the empty-cavity photon decay rate and $N
\gamma_e$ is a size-enhanced fluorescence rate. $\hat J_-$ is related to the dipole operator by $\hat \mu = \sqrt{N}(\hat J_-+\hat J_+) $, with $\hat J_+ = [\hat J_-]^\dagger$.


Leakage photoluminescence (LPL) is defined as $ S_{\rm LPL}(\omega)=\sum_j\rho_j\,  S_{\hat a}^{(j)}(\omega)$, where $S_{\hat a}^{(j)}(\omega)$ is given by Eq. (\ref{eq:S omega}) with $\hat O = \hat a$. In our simulations we use a stationary polariton distribution of the form $\rho_j=1/Md_j$, where $d_j$ is the degeneracy of the $j$-th polariton energy and $M=\sum_j d_j$ is the number of states considered. Leakage photoluminescence involves dissipative transitions in which the material is projected onto a state $\ket{\epsilon_i}$ in the ground manifold  after the cavity photon in a polariton state $\ket{\epsilon_j}$ decays through the mirrors. We can represent such process by the mapping
\begin{equation}\label{eq:PL map}
\hat a\ket{\epsilon_j}\rightarrow \ket{\epsilon_i}+\hbar\omega,
\end{equation}
where $\omega = \omega_j-\nu\,\omega_{\rm v}$ is the emitted light frequency, and $\nu$ is the number of vibrational quanta in  state $\ket{\epsilon_i}$. For $\nu\geq 1$, the emitted photon has an energy lower than its parent polariton, a fact that is  ignored in previous theories of organic microcavities \cite{Agranovich2003,Litinskaya2004,Litinskaya2006,Michetti2008,Mazza2009,Cwik2016}.  We show below that photon leakage transitions into states with up to $\nu=2$ vibrational quanta are important to interpret experiments.

The conventional absorption spectra $A(\omega)$ is obtained by introducing a laser driving term of the form $\hat V_{\rm p}(t) = \Omega_p\left(\hat a^\dagger\,{\rm e}^{-i\omega t}+\hat a\,{\rm e}^{i\omega t}\right)$ to the Holstein-Tavis-Cummings model,  $\Omega_p$ being a weak pumping strength and $\omega$ the driving frequency. Solving for the polariton population $\rho_j$ to second order in $\Omega_p$ for a system initially in the absolute ground state $\ket{G}$, setting $\hat O = \sqrt{N}\hat J_-$ in Eq. (\ref{eq:S omega}), and integrating over all dipole emission frequencies, gives the expression \cite{Herrera2016-PRA}
  \begin{equation} \label{eq:absorption}
 A(\omega_p) =\pi |\Omega_p|^2\sum_j\frac{|\langle G|\hat a|\epsilon_j\rangle|^2 (\kappa_{Gj}/\Gamma_j)}{(\omega_p-\omega_{jG})^2+\kappa_{Gj}^2}F_j,
 \end{equation}
where $\kappa_{Gj}$ is the decay rate for the coherence $\rho_{Gj} \equiv \bra{G}\hat \rho\ket{\epsilon_j}$, which we can allow to account for non-radiative relaxation processes \cite{Herrera2016-PRA}, and  $F_j = N\sum_i|\langle\epsilon_i|\hat J_-| \epsilon_j\rangle|^2$ is the total dipole emission strength of the $j$-th polariton. This expression shows that if $F_j = 0$ for a polariton eigenstate $\ket{\epsilon_j}$, there is no resonant absorption at that polariton frequency either.  In other words, polaritons that fluorescence poorly into bound modes of the nanostructure, cannot attenuate the reflected and transmitted fields efficiently.
 
In order to illustrate our theory, we show in Fig. \ref{fig:experiment} the simulated absorption and leakage photoluminescence (LPL) spectra for a system with $N=20$ emitters. This ensemble size is representative of the thermodynamic limit \cite{Zeb2016}. We set $\sqrt{N}\Omega/\omega_{\rm v}=2.4$, which is the critical coupling for the formation of a dark vibronic polariton $\ket{X}$ with zero energy $\omega_X = 0$ for $\lambda^2=1$. Fig. \ref{fig:experiment}(a) shows that  conventional absorption (dashed line) of the $\ket{X}$ state is negligible due to its small photonic component $\langle G|\hat a|X\rangle$, while bound absorption (solid line) of this state  is exactly zero because $\mu_X = 0$ \cite{Herrera2016-PRA}. The system also supports a second non-degenerate vibronic polariton $\ket{Xb} $ at frequency $\omega_j\approx 0.4\,\omega_{\rm v}$, which has a weak transition dipole moment.  Such state is weakly bright in both types of absorption measurements. On the contrary, there is no absorption from any of the multiple $Y$-type dark vibronic polaritons $\ket{Y_j}$ that exist in the frequency region between the lower and upper polariton peaks. This happens because for polaritons of the $Y$ type we have $\langle G|\hat a|Y_j \rangle = \langle G|\hat \mu|Y_j \rangle =0$.

 In Fig. \ref{fig:experiment}(b), we show the computed LPL spectra. Our results qualitatively reproduce  the experimental photoluminescence spectra obtained by Hobson {\it et al.} \cite{Hobson2002}, shown in Fig. \ref{fig:experiment}(c), and are also consistent with more recent measurements \cite{Schwartz2013}. In order to obtain a good qualitative agreement with the experimental spectra, it is important to take into account the emission from the dark vibronic polariton states labelled {\it X}, {\it Xb}, {\it Ya} and {\it Yb} in Fig. \ref{fig:experiment}(a) (inset) that leave the system with one quantum of vibration ($\nu=1$) in the ground manifold. Such emission enhances to the photoluminescence intensity near the lower polariton frequency, and blue shifts the peak maximum in that frequency region relative to the absorption peak. For the parameters in Fig. \ref{fig:experiment}, we obtain a blue shift  $\delta_{\rm LP}\approx 20$ meV for $\omega_{\rm v}= 180$ meV (vinyl stretch \cite{Spano2010}), which is consistent with recent measurements \cite{Schwartz2013}.

In order to quantify the importance of emission from dark vibronic polaritons of the $Y$ type, we show in Fig. \ref{fig:experiment}(d) the PL intensity $I_{\rm LP}$ collected at the frequency of the lower polariton state $\omega_{\rm LP}$, as a function of the center frequency $\omega_p$ of a Gaussian polariton population distribution with standard deviation $\sigma_p=0.5\,\omega_{\rm v}$. Such polariton distributions can be produced by a laser pulse \cite{Schwartz2013}. The figure shows that the emission peak at $\omega_{\rm LP}$ is greatly enhanced by {\it direct} photon leakage transitions from dark vibronic polariton states of the $Y$ type that occur at frequencies near the bare molecular resonance and also close to the upper polariton frequency.  Enhanced PL emission at the lower polariton frequency \cite{Ballarini2014} is thus not necessarily due to non-radiative relaxation on picosecond timescales \cite{Agranovich2003,Litinskaya2004,Litinskaya2006,Michetti2008,Mazza2009}. For metallic cavities with photon decay times $1/\kappa\sim 10$ fs \cite{Hobson2002, Schwartz2013,Coles2011,Virgili2011,George2015-farad}, sub-picosecond leakage emission from $Y$-type dark vibronic polaritons can already explain the observed emission enhancements.


 In summary, we have developed a theoretical framework based on the Holstein-Tavis-Cummings model \cite{Cwik2014,Spano2015,Herrera2016} to describe the spectroscopy of organic microcavities. The model provides a consistent microscopic interpretation for several observed features in the absorption and emission spectra of these systems for the first time. We do this by introducing dark vibronic polaritons, a class of light-matter excitations that in general involve an admixture of multiple material vibronic transitions with single photon states of the cavity that are dressed by purely vibrational material excitations.  For model parameters consistent with available experimental data, we show that emission from dark vibronic polaritons can appear in the vicinity of the bare molecular absorption frequency. Our theory does not exclude the possibility of residual emission from bare molecular states that are not strongly coupled to the cavity. However, since the absorption and emission spectra are found to strongly depend on the ratio between the Rabi coupling and the intramolecular vibration frequency, we expect the signals from weakly coupled emitters to behave differently from those of dark vibronic polaritons \cite{Herrera2016-PRA}.
 
Our developed theory can be further extended to treat microcavities in the ultrastrong coupling regime \cite{Schwartz2011,Kena-Cohen2013,Gambino2015} that have arbitrary mode dispersion, as well as materials with excitonic couplings, strong disorder or non-Markovian reservoirs. Understanding the ultrafast dynamics of organic microcavities may lead to the development of novel nonlinear optical devices \cite{Herrera2014,kowalewski2016cavity}, chemical reactors \cite{Hutchison:2012,Herrera2016}, and optoelectronic devices \cite{Feist2015,Schachenmayer2015,Orgiu2015,yuen2016} that are  enhanced by quantum optics.

\acknowledgements
We thank Bill Barnes for providing  the photoluminescence data from Ref. \cite{Hobson2002}, and Marina Litinskaya for technical discussions. F. Herrera is supported by CONICYT through  PAI N$^{\rm o}$ 79140030 and Fondecyt Iniciaci\'{o}n N$^{\rm o}$ 11140158. F. C. Spano is supported by the NSF, Grant N$^{\rm o}$ DMR-1505437.

\bibliographystyle{unsrt}
\bibliography{DVP}

\end{document}